# An ECC-based Fault Tolerance Approach for DNNs

Mohsen Raji[1], Mohammad Zaree[2], Kimia Soroush[3]
[1,2,3]School of Electrical and Computer Engineering, Shiraz University, Shiraz, Iran

*Abstract*—Deep Neural Network (DNN) has achieve great success in solving a wide range of machine learning problems. Recently, they have been deployed in datacenters (potentially for business-critical or industrial applications) and safety-critical systems such as self-driving cars. So, their correct functionality in the presence of potential bit-flip errors on DNN parameters stored in memories plays the key role in their applicability in safety-critical applications. In this paper, a fault tolerance approach based on Error Correcting Codes (ECC), called SPW, is proposed to ensure the correct functionality of DNNs in the presence of bit-flip faults. In the proposed approach, error occurrence is detected by the stored ECC and then, it is correct in case of a single-bit error or the weight is completely set to zero (i.e. masked) otherwise. A statistical fault injection campaign is proposed and utilized to investigate the efficacy of the proposed approach. The experimental results show that the accuracy of the DNN increases by more than 300% in the presence with Bit Error Rate of $10^{-1}$ in comparison to the case where ECC technique is applied, in expense of just 47.5% area overhead.

## I. Introduction

Deep Neural Networks (DNNs) are widely used in variety of applications such as object recognition from image or video input [1], [2] and natural language processing [3]. Their significant achievements have motivated their deployment in safety-critical applications such as military applications like Unmanned Aerial Vehicles (UAVs), space exploration [4] and autonomous driving [5]. However, any malfunctioning of DNNs used in such application (e.g. the navigation system of an autonomous vehicle) may lead to a catastrophic disaster.

Soft errors are known as a major source of faults that may lead to temporary malfunctions during the operation of a computing system running a DNN. These errors, typically caused by cosmic rays or alpha particles, may result in single-bit flips in memory cells storing the DNN parameters (i.e. weights and biases), which subsequently may lead to incorrect predictions or classifications [14].

There are some mitigating approaches to enhance the resilience of DNNs against soft errors. Hardware redundancy offers a layer of protection by utilizing backup components to isolate and neutralize the effects of faulty units. Software-based solutions, such as error-aware training [15] and resilient architectures, are also being explored to mitigate the impact of soft errors on DNNs. However, hardware or software redundancy-based often come with trade-offs such as increased complexity, reduced performance, or higher costs [16].

One effective strategy to improve the soft error resilience of DNNs is the implementation of Error Detection and Correction Codes (ECC). ECC mechanisms can identify and, in some cases, rectify soft errors before they cause significant disruptions, in expense of some memory overheads. These approaches aim to design models capable of tolerating or recovering from occasional bit flips [17]. In comparison to hardware or software redundancy, ECC operates at a fundamental level, providing a means to detect and correct errors in data without the need for additional hardware or significant modifications to software architecture.

In this paper, a fault tolerance approach based on ECC, called SPW, is proposed to ensure the correct functionality of DNNs in the presence of bit-flip faults. In the proposed approach, error occurrence is detected by the stored ECC and then, it is correct in case of a single-bit error or the weight is completely set to zero (i.e. masked) otherwise. Single Error Correction Double Error Detection (SECDED) is used as the ECC which is capable of correcting single bit flips and detecting double bit flips.

In order to investigate the efficacy of the proposed method, a set of extensive experiments based on fault injection (FI) approach is conducted. To ensure that fault injection process is complete, we present a statistically complete FI tool. We inject the faults with different fault rates and noticed that there is a point in fault injection which as fault rate increased, the accuracy does not change and shows the network is completely operating randomly, called as *saturation point*. The experimental results show that the accuracy of the DNN increases by more than 300% in the presence with Bit Error Rate (BER) of $10^{-1}$ in comparison to the case where ECC technique is applied, in expense of just 47.5% area overhead.

The remainder of this paper is organized as follows. Section II provide background information about neural networks and CNNs. Section III presents different works which has been done in reliability of neural networks in different contexts. Section IV explains the fault injection tool and the way which we use it. Section V presents the proposed fault tolerant technique for CNNs. Section VI shows different results related to the proposed approach and finally Section VII concludes the paper.

## II. Background

This section provide background information required to follow the rest of this paper.

### A. DNN

Fig. 1 shows the overall structure of a DNN. DNNs consist of one input layer, one output layers, and multiple hidden layers in between. Each layer contains neurons that receive inputs, process them through weights and biases, and pass the results to the next layer. The output layer produces the final prediction based on the processed inputs from the hidden layers.

Weights are numerical values that adjust the influence of incoming data in a neural network, determining how much each input affects the output. During training, weights are fine-tuned to improve predictions. In DNNs, weights are critical for learning and decision-making, so any modifications, like bit flip errors, can significantly impact accuracy and performance. This highlights the importance of protecting weights to ensure proper network functionality.

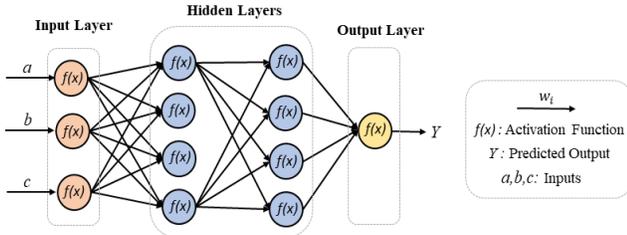

Fig. 1. DNN Optimization and Learning

### B. SECDED

Error Correction Codes (ECC) ensure data integrity in memory by detecting and correcting bit upsets caused by soft errors, such as single-bit flips due to cosmic rays or radiation. ECC adds extra bits to monitor data consistency. A common ECC implementation is Single Error Correction Double Error Detection (SECDED) using Hamming codes. SECDED detects two errors and corrects one by identifying the flipped bit through parity bits or other techniques, ensuring data reliability.

The general algorithm to generate a SECDED code for any number of bit is as follows:
1) Arrange bits position starting from 1: bit 1, 2, 3, 4, etc.
2) All bit positions powers of two ($2^n$) are parity bits generated by applying XOR operation on data bits
3) All other bit positions are data bits in sequence
4) Each data bit is covered with parity bits of its binary form position

Fig. 2 shows the SECDED code generated for a 26-bit data.

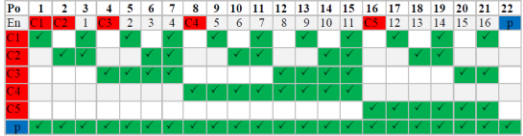

Fig. 2. Computing parity bits for a 16-bit data

During storage the bits, the parity bits are computed and stored in memory. During retrieving the data bits, the parity bits are re-computed and checked against the stored parity bits. If a single discrepancy is found, it indicates a single-bit error. The location of the parity bit that failed to match is used to identify and correct the single-bit error. However, if two parity bits fail, indicating a double-bit error, it can be detected but cannot be corrected.

In summary, SECDED codes represent a foundational approach to error detection and correction in storing data such as DNN parameters, offering a practical solution for ensuring data integrity in the presence of occasional errors.

## III. RELATED WORKS

This section presents the research works focused on enhancing the soft error resilience of DNNs.

In [6] a recurrent relation is proposed for neuron resiliency to faults. Based on this relation, neurons categorized in high and low resiliency to faults. As a fault tolerant technique, neurons that have high resiliency to faults are mapped to Protected Processing Elements (PPE) and other neurons are mapped to conventional PEs.

Resiliency of a FPGA to faults is evaluated by three types of injecting faults including radiation, laser and emulation in [7]. This work shows that each type has its own advantages and disadvantages. For example, laser fault injection can inject faults into each block of FPGA.

Injecting one fault into bit-stream of FPGA when loading on it for a CNN that its parameters has been implemented just by one-bit has no impressive effect while multiple faults have large effects [8].

In [9] reliability of neural networks against radiation faults is assessed and then Triple Modular Redundancy (TMR) and selective TMR is applied to be fault tolerant against faults reaches to output (Critical Errors).

Quantization is effective for radiation faults. In [10] CNN is changed to HNN (Hybrid Neural Network) and then is showed that radiation sensitivity is reduced because the cross section of the design is reduced. HNN is the CNN with one-bit parameters for convolution layers.

In [11] a co-design tool (which is called Minerva) is introduced in algorithm, architecture and circuit level of neural networks in order to reduce the consumption power. Therefor the fault rate is increased. To prevent errors because of power reduction, A RASOR Double Sampling is used for fault detection and Bit Masking (in which the faulty bit is set to the sign bit) and Word Masking (in which all bits are set to zero) is applied for fault correction in circuit level. By this method power consumption is less than 8.1 times from the original design. In [12], RASOR Double Sampling is used as fault detection mechanism but a variant of Bit Masking is applied which is different from the Bit masking method in [11]. In this work if fault occurs in sign bit, it is replaced with most significant bit and if fault occurs in most significant bit, that word is set to zero. If fault occurs in other bits, that bit is replaced with sign bit. This method is the combination of bit masking and word masking that reaches more fault tolerance in comparison with [11]. Bit flip fault model especially flipping from 0 to 1 has more effect than flipping from 1 to 0 [13]. This work suffers from the high area overhead and thus, is not efficient in real applications.

## IV. THE PROPOSED FAULT TOLERANCE APPROACH: SPW

In this section, the proposed fault tolerant technique for protecting the DNN parameters (weights and biases), named as SPW, is presented.

### A. Error Resilience Capability of the Proposed Approach

The ECC used in the proposed approach is SECDED code which can correct a single error and detect if two errors occurs. However, in case of two errors, SECDED Code has no correction capability. So, the proposed approach takes advantages of word masking technique to prevent the error propagation in DNNs when two errors occurs. So, the error resilience property of the proposed approach is as follows:
1. If a single error occurs in a parameter, the proposed approach detects and corrects it using error recovery capability provided by SECDED code.
2. If two error occur in a parameter, the errors are detected by detection capability provided by SECDED code and then, all the bits of the erroneous parameter are set to zero, preventing to propagate the error in the rest of the DNN.

Fig. 3 shows the block diagram of the proposed the proposed approach applied in neuron-level to provide a soft error resilient neuron. Each soft error resilient neuron has different input/output and components, including:

1. $Input$: the input of a neuron.
2. $Par$: the weights or biases of the network.
3. $Parity$: the parity bit of the parameters.
4. $Output$: the output of neuron.
5. $Arithmetic$ unit: the function of a neuron, which can be convolution operations or activation functions, and so on.
6. $SPW$ unit: the unit added to a neuron that is responsible for error detection and correction.
7. $CPar$: the output of $SPW$ unit which is the input of the $Arithmetic$ unit.

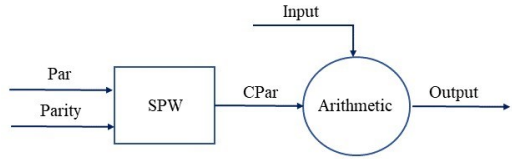

Fig. 1. Fault Tolerant Neuron

The architectural block diagram of the proposed fault tolerant unit is shown in Fig. 2. This unit corrects if a single error occurs in a parameter and resets it to zero if two errors occur.

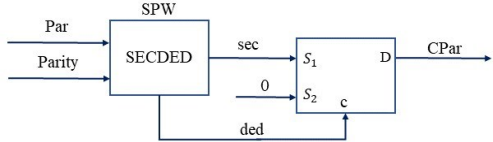

Fig. 2. Architecture of the SPW unit

The $sec$ signal, which is an N-bit output (the value of N is equal to the number of bits needed to display a parameter), is the corrected value of the parameter, and $signal$ is a one-bit output that indicates whether two faults have occurred in a parameter or not. The SECDED unit consists of four important units as shown in Fig. 3.

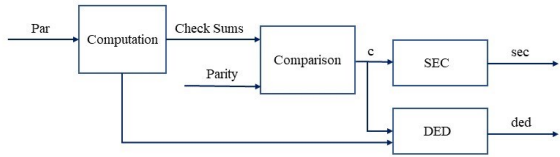

Fig. 3. Different Parts of SPW unit

1. Computation
   In this unit, the operation of calculating the parity bit of the parameter entered into the neuron is performed. This unit consists of a series of logic gates that are calculated according to the for 16-bit data.
2. Comparison
   In this unit, the location of the single fault is determined (if exists). At the first the formula 1 is computed and then XORed with the parity (the input of this unit).
$$Par_{parameter} = C_1 C_2 \ldots C_l$$
$$C = Par_{parameter} \oplus parity$$
3. SEC
   If a single fault occurred, corresponding bit is flipped. P is the Xor of all bits, including P itself, which is a one-bit output obtained from the Computation unit.
4. DED
   Table 1 shows different situations of fault occurrence according to the values of C and P; i.e. it determines whether two bits of fault occurred in a parameter and what the output signal of this module (DED signal) should take.

TABLE 1. DETECTION OF FAULT TYPE

|  | $P = 0$ | $P = 1$ |
|---|---|---|
| $C = 0$ | No Fault | P is faulty |
| $C \neq 0$ | Double Fault | Single Fault |

This method has overheads that must be addressed so that the user is aware of them to use this method. This method generally has two different parts from the overhead perspective, which are:

- Parity bits: The parity bits required for this method is the same as the parity bits required for the SECDED method.
- SPW unit: In this method, one the proposed approach unit is added to each neuron which has some logical gates.

## V. STATISTICAL FAULT INJECTION TOOL

It is a common method to inject defects in the different sections randomly which has several major drawbacks including:

1. Very large space of points where faults must be injected
2. Inability to provide a statistical guarantee that the injection method is complete

In order to get rid of the mentioned problems, a Statistical Fault Injection (SFI) is used as an injector of faults in the network parameters. Therefore, the following steps must be taken to inject the faults to claim that SFI has been used correctly.

1. The network is trained in a fault-free mode to achieve the weights and bias that lead to maximum accuracy. The network parameters obtained at this stage must be in the form of a probabilistic distribution. Usually the distribution considered for this is a normal distribution $N(\mu, \sigma)$. $\mu$ represents the mean and $\sigma$ represents the standard deviation. It should be noted that if $\sigma$ equals to zero, it means that the random variable is practically converted to a fixed number with the value of $\mu$.
2. A model for applying transient errors must be specified. For this problem, a Bernoulli distribution corresponding to each bit of the network parameters is considered with probability p. So the formula for producing faulty parameters is given by
$$Par_{faulty} = Par_{correct} \oplus e$$
3. At the end of the second stage, an output distribution

is obtained for each neuron, which forms a DBN.

4. In order to obtain the accuracy of the network in the faulty mode, it is necessary to give the input to network and gather the output using MCMC algorithms by different values of p.

To better illustrate steps 1 to 4, the pseudo code for it, is given in Algorithm 4.

```
Algorithm 1
1: Train network in a fault-free scheme to get weights (W) and biases (B)
2: distribution_list = []
3: p = Any arbitrary number between [0, 1]
4: b = p^M where M = Any arbitrary sufficient natural number
5: accuracy_previous = 0
6: for iteration from 1 to M do
7:   Compute error distribution for all weights (E_w) and biases (E_b) by ComputeError(p)
8:   W_faulty = W + E_w
9:   B_faulty = B + E_b
10:  correct_predictions = 0
11:  for element in D do
12:    output = Inference(W_faulty, B_faulty, element)
13:    if output == Golden_run_output then
14:      correct_predictions = correct_predictions + 1
15:    end if
16:  end for
17:  N = Number of elements in D
18:  accuracy_current = correct_predictions / N
19:  Compute a random uniform number between [0, 1].
       Then: random_number = rand() / MAX_RAND
20:  if random_number < min(1, accuracy_current / accuracy_previous) then
21:    Add accuracy_current to distribution_list
22:    accuracy_previous = accuracy_current
23:  else
24:    Add accuracy_previous to distribution_list
25:  end if
26: end for
```

Algorithm 4. Pseudo code for SFI

In which Inference is the conclusion phase of the neural network to which the input (element) is given and the output is taken. M is a natural number that indicates the number of times that the network parameters must be faulty. This number should be large enough to report a good estimate of output accuracy. D is a set of inputs that must be given as input to the network. The pseudo code of the Compute Error function is also given in Algorithm 5. The rand function generates three random numbers and returns the smallest of them.

```
1: for each bit in P do
2:   if p ≥ rand() / MAX_RAND then
3:     Set the corresponding bit in E_g or E_b to 1
4:   end if
5: end for
```

Algorithm 5. Pseudo code for *rand* function

As shown in Algorithm 4, this algorithm is similar to the metropolis algorithm. In this case, the accuracy of the network is the same as the objective function and the element that must be (consciously) sampled is network parameters. The accuracy of the network in this issue is equivalent to (x) and the network parameters are equivalent to input x in the metropolis algorithm. Note that presence in a state of network parameters depends only on a state before it. The kernel function used here is a Bernoulli distribution due to the nature of the problem rather than a normal distribution.

$$\text{Par}_{\text{current}} = \text{Ber}(\text{Par}_{\text{previous}}, p)$$

The Ber(.) function means that each bit of the previous network parameters ($\text{Par}_{\text{previous}}$) may be flipped with a probability of p. The output of this function is the current parameters ($\text{Par}_{\text{current}}$). It should also be noted that the kernel function in this case is a symmetric function because:

$$Q(\text{Par}_{\text{previous}}|\text{Par}_{\text{current}}) = Q(\text{Par}_{\text{current}}|\text{Par}_{\text{previous}}) = (1-p)^n * p^m$$

As shown in relation 3, to go from $\text{Par}_{\text{previous}}$ state to $\text{Par}_{\text{current}}$ and vice versa, n bits must not be flipped and m bits must be flipped. Therefore, the probability acceptance function will be as follows:

$$A = \min\left(1, \frac{Q(\text{Par}_{\text{previous}}|\text{Par}_{\text{current}}) * \text{accuracy}_{\text{currect}}}{Q(\text{Par}_{\text{current}}|\text{Par}_{\text{previous}}) * \text{accuracy}_{\text{previous}}}\right)$$
$$= \min\left(1, \frac{\text{accuracy}_{\text{currect}}}{\text{accuracy}_{\text{previous}}}\right)$$

Therefore, according to the Metropolis algorithm, the current accuracy may be rejected and the previous value accepted as the output.

## VI. EXPERIMENTAL RESULTS

This chapter empirically examines the claim made in the previous chapter. This chapter consists of several sub sections, each of which is done with specific conditions that are discussed below.

### A. Experiment Setup

In the experiments, the LeNet neural network is used to detect MNIST dataset numbers. General specifications of LeNet neural network is in Table 2. The network parameters are stored in 16-bit precision and the operations related to the parameters are performed as a fixed point number representation. The training and inference phases of the parameters of is performed in Python environment using cross platform.

**TABLE 2.** CHARACTERISTIC OF THE NEURAL NETWORK

| Type | Deep Convolutional Neural Network |
|---|---|
| Topology | 7L(2 Convolution layer (8 and 16) filters respectively), 2 Max pooling layer by stride(2,2), 3 Fully Connected layers) |
| Number of Weights per Layer | (128,512,69120,10080,840) = 80680 |
| Activation Function | Relu |

Transient faults in various network parameters are injected using the SFI tool. According to the algorithm in Algorithm 4, the network parameters become faulty for 100 times, and each time 10,000 images are given to the network to obtain accuracy. A boxplot is then drawn for the numbers obtained. Fault injection is done in two different cases, in the first case, each bit of parameter can be faulty (no limitation on the location and number of bit-flip fault on a parameter) and in second case maximum of two faults in one parameter may be injected (limitation on the number of faults). In each part, the faults are injected in three different modes, which are:

1. All layers
2. Only convolution layers
3. Only FC layers

### B. Injecting faults on the network without fault tolerant technique

In this part, after injecting the faults, input is given to the network and no fault tolerant technique has been applied on the network.

The effect of fault on the FC layers is greater than the convolution layers.

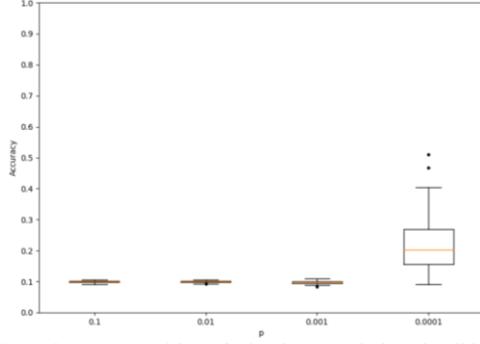

Fig. 7. Network accuracy without fault tolerant technique in all layers. p is the probability of bit flipping in each bit of each parameter.

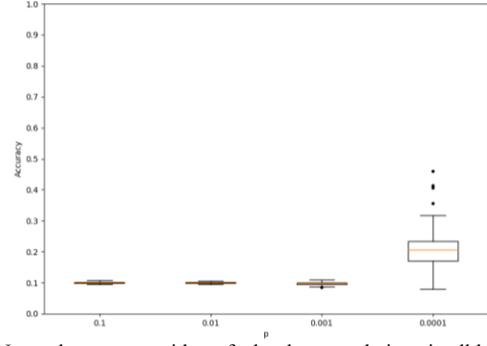

Fig. 8. Network accuracy without fault tolerant technique in all layers with limitation of injecting at most two faults in each parameter. $p$ is the probability of bit flipping in each bit of each parameter.

As can be seen in Table 3, the average network accuracy did not change significantly at p = 0.1 and p = 0.01. Therefore, it can be concluded that if the fault rate exceeds a certain level, the network operates completely as random guess, which we call network saturation in injecting faults.

*C. Injecting faults on the network with ECC technique*

In this experiment, after injecting the faults, input is given to the network in which only SECDED ECC technique is applied on the network.

Fig. 9 and Fig. 10 show the obtained results for FI experiment. As the result show, the impact of FI on FC layers is more significant comparing to convolution layers.

*D. Injecting faults on the network with the proposed approach technique*

In this experiment, the proposed approach technique is applied to the network in different modes. The obtained FI results are shown in Fig. 11 and Fig. 12.

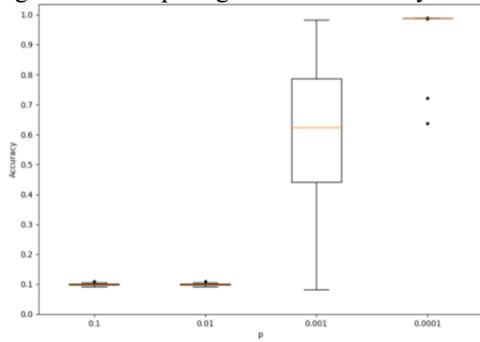

Fig. 9. Network accuracy with ECC technique in all layers. p is the probability of bit flipping in each bit of each parameter.

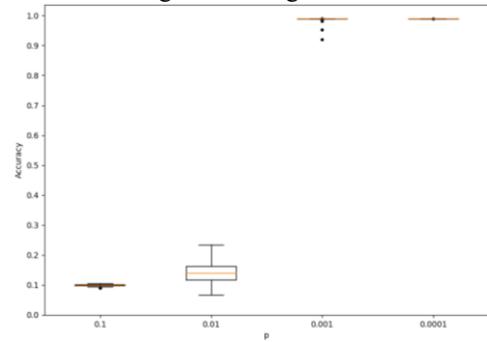

Fig. 11. Network accuracy with the proposed approach in all layers. p is the probability of bit flipping in each bit of each parameter.

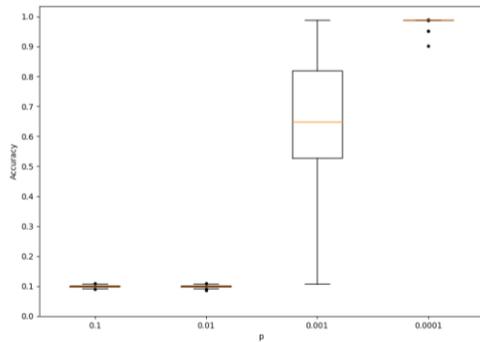

Fig. 10. Network accuracy with ECC technique in all layers with limitation of injecting at most two faults in each parameter. p is the probability of bit flipping in each bit of each parameter.

The summary of results in different situations has been reported in Table 3.

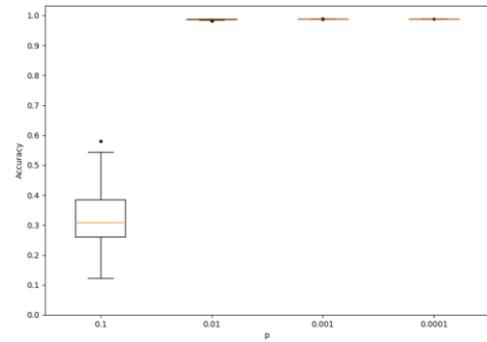

Fig. 12. Network accuracy with the proposed approach in all layers with limitation of injecting at most two faults in each parameter. p is the probability of bit flipping in each bit of each parameter.

Median accuracy of results are reported in Table 5. It is notable that in p = 0.1, the network does not saturate when there is the limit of injecting at most 2 faults in each parameter. This part shows that the proposed approach is a good technique over ECC. And also the accuracy of the network is significantly high at p = 0.01 compared to ECC.

TABLE 3. MEDIAN OF ACCURACY FOR DIFFERENT INJECTING FAULT RATES (P) WITHOUT FAULT TOLERANT TECHNIQUE.

| p | Accuracy | | | | | | | |
|---|---|---|---|---|---|---|---|---|
| | 0.1 | | 0.01 | | 0.001 | | 0.0001 | |
| Limitation in Faults | No | 2 | No | 2 | No | 2 | No | 2 |
| All | 0.0997 | 0.0995 | 0.0994 | 0.1 | 0.0984 | 0.0978 | 0.2326 | 0.2137 |
| FC | 0.1005 | 0.1001 | 0.0999 | 0.0996 | 0.0958 | 0.0958 | 0.2232 | 0.2173 |
| Convolution | 0.1 | 0.0996 | 0.1 | 0.0999 | 0.5734 | 0.5549 | 0.9889 | 0.9885 |

TABLE 4. MEDIAN OF ACCURACY FOR DIFFERENT INJECTING FAULT RATES (P) WITH ECC TECHNIQUE

| p | Accuracy | | | | | | | |
|---|---|---|---|---|---|---|---|---|
| | 0.1 | | 0.01 | | 0.001 | | 0.0001 | |
| Limitation in Faults | No | 2 | No | 2 | No | 2 | No | 2 |
| All | 0.1 | 0.1004 | 0.0997 | 0.0992 | 0.6649 | 0.7175 | 0.9826 | 0.9872 |
| FC | 0.1 | 0.1003 | 0.0984 | 0.0967 | 0.6854 | 0.7405 | 0.9849 | 0.9881 |
| Convolution | 0.0990 | 0.0995 | 0.3976 | 0.4751 | 0.9833 | 0.9888 | 0.9889 | 0.9889 |

TABLE 5. MEDIAN OF ACCURACY FOR DIFFERENT INJECTING FAULT RATES (P) WITH THE PROPOSED APPROACH TECHNIQUE

| p | Accuracy | | | | | | | |
|---|---|---|---|---|---|---|---|---|
| | 0.1 | | 0.01 | | 0.001 | | 0.0001 | |
| Limitation in Faults | No | 2 | No | 2 | No | 2 | No | 2 |
| All | 0.992 | 0.3086 | 0.1397 | 0.9873 | 0.9888 | 0.9889 | 0.9889 | 0.9889 |
| FC | 0.1001 | 0.7687 | 0.1436 | 0.9881 | 0.9888 | 0.9889 | 0.9889 | 0.9889 |
| Convolution | 0.0996 | 0.5022 | 0.9851 | 0.9879 | 0.9889 | 0.9889 | 0.9889 | 0.9889 |

TABLE 6. COMPARISON OF THE PROPOED TECHNIQUES WITH THE STATE-OF-THE-ART TECHENIQUES IN TERMS OF ACCURACY

| Technique | Accuracy | Number of Injected Faults | Hardware Redundancy | Fault Detection | Fault Correction |
|---|---|---|---|---|---|
| No Technique | 0.2459 | 130 | – | - | - |
| TMR [10] | 0.9889 | 1 | 200% | Deterministic | Deterministic |
| WM [12] | 0.6433 | 16 | 2.6% | Probabilistic | Probabilistic |
| ECC | 0.9826 | 171 | 47% | Deterministic | Deterministic |
| ECC+WM | 0.9889 | 181 | 47.5% | Deterministic | Probabilistic |

*E. Discussion and Comparison with the State-of-the-art*

According to the results obtained from ECC and the proposed approach methods, it can be concluded that when a fault tolerant method is applied in all cases where the fault rate is equal to 0.0001, the average accuracy of the network is significantly larger than the non-fault tolerant technique and close to the fault-free value (according to Table 3, 4, and 5)

The median accuracy of network in the proposed approach is higher than in the ECC method in most cases (according to

The advantage of the proposed approach over ECC method is especially obvious when there is a maximum injection limit of two faults in each parameter.

The accuracy of the network in the proposed approach is 1.53 times the accuracy of the network in [12] method, which shows the superiority of the proposed approach over the method [12]. It should be noted that the closest case to the method [12] in this article is to use p = 0.0001, which can be compared.

The method [12], although a very inexpensive method, has two main problems.
1. If the number of fault injections exceeds the number mentioned in their article, the accuracy of the network decreases sharply and reaches saturation. Therefore, this method cannot be used in applications which there is a significant drawback.
2. Although this method has very little overhead in fault detection (2.6% of the total resources of a design without applying technique), but fault detection is probabilistic and this makes it impossible to use this method insensitive and critical applications.

VII. CONCLUSION

DNNs have achieved significant success in addressing various machine learning challenges. Their deployment in data centers and safety-critical systems, such as autonomous vehicles, underscores the importance of ensuring their correct operation despite potential hardware failures. This paper proposes a fault-tolerant approach using ECC to maintain DNN functionality even when bit-flip errors occur in memory-stored parameters. The proposed method detects errors through ECC and either corrects them for single-bit mistakes or masks the entire weight if necessary. To evaluate its effectiveness, the authors conducted a statistical fault injection study. The results indicate that DNN accuracy improves by over 300% at a $10^{-1}$ Bit Error Rate compared to using ECC alone, albeit at the cost of approximately 47.5% increased area usage.

It should be noted that in the continuation of this work, improvements can be added to this method in order to reduce its overhead. Also, the selection of fault-tolerant neurons can be more intelligent and based on the selection of neurons that have a greater impact on network output, rather than being layered.

REFERENCES

[1] A. Krizhevsky, I. Sutskever, and G. E. Hinton, Imagenet classification with deep convolutional neural networks, in Advances in neural information processing systems, 2012, pp. 10971105.

[2] A. Karpathy, G. Toderici, S. Shetty, T. Leung, R. Sukthankar, and L. FeiFei, Large-scale video classification with convolutional neural networks, in Proceedings of the IEEE conference on Computer Vision and Pattern Recognition, 2014, pp. 17251732.

[3] Y. Kim, Convolutional neural networks for sentence classification, arXiv preprint arXiv:1408.5882, 2014.

[4] G. R. Allen, S. Vartanian, F. Irom, L. Z. Scheick, P. Adell, M. D. OConnor, and S. M. Guertin, 2017 compendium of recent test results of single event effects conducted by the jet propulsion laboratorys radiation , effects group, in 2017 IEEE Radiation Effects Data Workshop (REDW), July 2017, pp. 715.

[5] V. Neagoe, A. Ciotec, and A. Brar, A concurrent neural network approach to pedestrian detection in thermal imagery, in 2012 9th International Conference on Communications (COMM), June 2012, pp. 133136.

[6] Schorn, Christoph & Guntoro, Andre & Ascheid, Gerd. (2018). Accurate neuron resilience prediction for a flexible reliability management in neural network accelerators. 979-984. 10.23919/DATE.2018.8342151.

[7] p. 10.1109/SBCCI.2018.8533235., Benevenuti, Fabio & Libano, Fabiano & Pouget, Vincent & Kastensmidt, Fernanda & Rech, P.. (2018). Comparative Analysis of Inference Errors in a Neural Network Implemented in SRAM-Based FPGA Induced by Neutron Irradiation and Fault Injection Methods. 1-6.

[8] Du, Boyang & Azimi, Sarah & De Sio, Corrado & Bozzoli, Ludovica & Sterpone, Luca. (2019). On the Reliability of Convolutional Neural Network Implementation on SRAM-based FPGA. 1-6. 10.1109/DFT.2019.8875362.

[9] Libano, F. & Wilson, B. & Anderson, J. & Wirthlin, M. & Cazzaniga, Carlo & Frost, Christopher & Rech, P.. (2018). Selective Hardening for Neural Networks in FPGAs. IEEE Transactions on Nuclear Science. PP. 1-1. 10.1109/TNS.2018.2884460.

[10] p. 10.1109/TNS.2020.2983662., Libano, F. & Wilson, B. & Wirthlin, M. & Rech, P. & Brunhaver, J.. (2020). Understanding the Impact of Quantization, Accuracy, and Radiation on the Reliability of Convolutional Neural Networks on FPGAs. IEEE Transactions on Nuclear Science. PP. 1-1.

[11] pp. Architecture News. 44. 267-278. 10.1145/3007787.3001165., Reagen, Brandon & Whatmough, Paul & Adolf, Robert & Rama, Saketh & Lee, Hyunkwang & Lee, Sae & Hernndez-Lobato, Jos & Wei, GuYeon & Brooks, David. (2016). Minerva: Enabling Low-Power, Highly-Accurate Deep Neural Network Accelerators. ACM SIGARCH Computer.

[12] Salami, Behzad & Unsal, Osman & Cristal, Adrian. (2018). On the Resilience of RTL NN Accelerators: Fault Characterization and Mitigation.

[13] Kim, Jae-San & Yang, Joon-Sung. (2019). DRIS-3: Deep Neural Network Reliability Improvement Scheme in 3D Die-Stacked Memory based on Fault Analysis. 1-6. 10.1145/3316781.3317805.

[14] Ibrahim, Younis, et al. "Soft errors in DNN accelerators: A comprehensive review." *Microelectronics Reliability* 115 (2020): 113969.

[15] Safari, Sepideh, et al. "A survey of fault-tolerance techniques for embedded systems from the perspective of power, energy, and thermal issues." *IEEE Access* 10 (2022): 12229-12251.

[16] Hoang, Le-Ha, Muhammad Abdullah Hanif, and Muhammad Shafique. "Ft-clipact: Resilience analysis of deep neural networks and improving their fault tolerance using clipped activation." *2020 Design, Automation & Test in Europe Conference & Exhibition (DATE)*. IEEE, 2020.

[17] Hanif, Muhammad Abdullah, and Muhammad Shafique. "Dependable deep learning: Towards cost-efficient resilience of deep neural network accelerators against soft errors and permanent faults." *2020 IEEE 26th International Symposium on On-Line Testing and Robust System Design (IOLTS)*. IEEE, 2020